\documentstyle[preprint,aps]{revtex} 
\renewcommand{\bbox}[1]{\mbox{\boldmath $#1$}} % corresponds to \vec          
\begin{document}                                     
\title{A Contracted Path Integral Solution of the Discrete Master Equation}
\author{Dirk Helbing}
\address{II. Institute for Theoretical Physics, University of
Stuttgart, 70550 Stuttgart, Germany}
\maketitle      
\begin{abstract}
A new representation of the exact time dependent solution
of the discrete master equation is derived. 
This representation can be considered
as contraction of the path integral solution of {\sc Haken}. 
It allows the calculation of the probability distribution of the 
occurence time for each path
and is suitable as basis of new computational solution methods.
\end{abstract}
\section*{Introduction}

The master equation \cite{Pau28,Gar85,Hak83,Hel92}        
plays a very important role in the description of
stochastically behaving systems. In statistical physics it arises in the
{\sc Markoff}ian limit \cite{Kenkre} of the {\em generalized master equation}
\cite{Kenkre}. The generalized master equation is obtained as a special
representation of the {\sc Nakajima-Zwanzig} equation, which results
by projection of the {\sc von Neumann} equation 
for the {\em statistical operator}
\cite{Fano} on the relevant variables of the 
considered system \cite{Zwa,FiSa}.
Applications of the master equation reach from nonequilibrium thermodynamics
\cite{Kamp,Haus,Hak}, 
over chemistry \cite{Kamp,Opp} and biology \cite{Lef} to the social
sciences \cite{WeHa83,We91,Hel93,Hel93a,Hel93b}. 
\par
In most cases a concrete master equation is only numerically soluble. For a
small number of possible system states 
computer programs for the solution
of a linear system of first order ordinary differential equations can be used.
However, in many situations {\em Monte Carlo simulations}
\cite{Bin} have to be applied, yielding only approximate results.
An alternative approach is the {\em path integral solution} 
discussed below. It provides an {\em exact} expression for the {\em time
dependent} solution even of the {\em generalized} master equation, for which
very few solution methods exist up to now. This expression is a suitable
starting point for the development of new {\em computational} solution methods.
It also allows the calculation of the probability that a system takes a 
`desired' or a `catastrophic' path, which is very important for technical
and other applications.

\section*{The path integral solution}

Let $p(x,t)$ with $0 \le p(x,t) \le 1$ and $\sum_x p(x,t) = 1$
denote the {\em probability} of the considered system to be in 
{\em state} $x$ at time $t$. Further, for $x' \ne x$ let
\begin{equation}
 w(x'|x;t) = \lim_{\Delta t \rightarrow 0} \frac{p(x',t+\Delta t|x,t)}{\Delta t}
\end{equation}
be the {\em transition rate} (i.e., the 
{\em transition probability} $p(x',t+\Delta t|x,t)$
per time unit $\Delta t$) from state $x$ to state $x'$. Then, the {\em master
equation} for the temporal development of the probability distribution
of the system reads
\begin{equation}
 \frac{dp(x,t)}{dt} = \sum_{x'(\ne x)} [ w(x|x';t)p(x',t) 
 - w(x'|x;t) p(x,t) ] \, .
\label{master}
\end{equation}
Having the limit $\Delta t \rightarrow 0$ in mind, this equation can be
written in the form
\begin{eqnarray}
 p(x,t+\Delta t) &=& p(x,t) +
 \Delta t \sum_{x'(\ne x)} w(x|x';t)p(x',t) 
 - \Delta t \sum_{x'(\ne x)} w(x'|x;t) p(x,t) \nonumber \\
 &=&  \sum_{x'} p(x,t+\Delta t|x',t) p(x',t)
\end{eqnarray}
with 
\begin{equation}
 p(x,t+\Delta t|x',t) = \delta_{xx'} 
 + \Delta t  \sum_{x^{\prime\prime}(\ne x)}
 [ w(x|x^{\prime\prime};t) \delta_{x'x^{\prime\prime}} ] 
 - \delta_{xx'} \Delta t \sum_{x^{\prime\prime}(\ne x)} 
 w(x^{\prime\prime} | x;t) \, ,
\end{equation}
which is also called a {\em propagator},
and the {\sc Kronecker} function
\begin{equation}
 \delta_{xx'} := \left\{
\begin{array}{ll}
1 & \mbox{if } x=x' \\
0 & \mbox{otherwise.}
\end{array} \right.
\end{equation}
The solution of the master equation (\ref{master}) is, therefore,
\begin{equation}
 p(x,t) = \lim_{N\rightarrow \infty} \sum_{x_{N-1}} \sum_{x_{N-2}} \dots
 \sum_{x_0} 
 \left[ \prod_{i=1}^N p(x_i,t_i|x_{i-1},t_{i-1}) \right]
 p(x_0,t_0) \, ,
\label{haken}
\end{equation}
where $x_N := x$, $t_i := t_0 + i\Delta t$, and $\Delta t := (t-t_0)/N$.
This representation is a consequence of the {\sc Chap\-man-Kol\-mo\-go\-rov} 
equation \cite{Gar85}, and it is the 
basis of {\sc Haken}'s approximate path integral solution
of the master equation \cite{Hakpath}. 
\par
In the following, we will restrict our considerations to time independent
transition rates $w(x'|x;t) \equiv w(x'|x)$ and derive a contracted
form of expression (\ref{haken}). In order to achieve this simplification
we utilize the fact that the system will normally not change the
state $x_i$ over many time periods $\Delta t$. 
That means, within the time interval
$\tau:= t - t_0$ the system will usually have changed its state $x_i$ a finite 
number of times only. Let $x_0, x_1, \dots, x_n$ be the sequence of states
the system shows between times $t_0$ and $t$. Then,
\begin{equation}
 {\cal C}_n := x_0 \rightarrow x_1 \rightarrow \dots \rightarrow x_n
\end{equation}
with $x_n := x$ shall be called the {\em path} of the system, and 
\begin{equation}
 p({\cal C}_n,\tau) \equiv p(x_0\rightarrow \dots \rightarrow x_n,\tau)
\end{equation}
shall denote the probability that the system takes the path ${\cal C}_n$
during the time interval $\tau$. This implies the relation
\begin{eqnarray}
 p(x,t) &=& p(x,t_0+ \tau) 
 = \sum_{n=0}^\infty \sum_{{\cal C}_n} p({\cal C}_n,\tau) \nonumber \\
&\equiv& \sum_{n=0}^\infty \; \sum_{x_{n-1} (\ne x_n)} \;
 \sum_{x_{n-2}(\ne x_{n-1})}
 \dots \sum_{x_0(\ne x_1)} 
 p(x_0 \rightarrow \dots \rightarrow x_n,\tau) \vphantom{\sum} \, .
\label{contraction}
\end{eqnarray}
The representation (\ref{contraction}) of the solution of the master equation
(\ref{master}) is the desired contraction of formula (\ref{haken}) and
has been suggested by {\sc Empacher} \cite{Nils}. In order to derive the
explicit form of the {\em path probabilities} $p({\cal C}_n,\tau)$, we will
extend an idea of {\sc Weidlich} \cite{Weid}. For this 
purpose we rewrite the master equation (\ref{master}) in vectorial form
\begin{equation}
 \frac{d\bbox{p}(t)}{dt} = (W - D ) \bbox{p}(t)
\label{master2}
\end{equation}
with
\begin{equation}
 \bbox{p}(t) := (p(1,t), \dots, p(x,t), \dots )^{\rm tr}
\end{equation}
and the matrices
\begin{eqnarray}
 W \equiv \Big(W_{xx'}\Big) &:=& \Big( w(x|x') \Big) \, , \nonumber \\
 D \equiv \Big( D_{xx'} \Big) &:=& \Big( w_x \delta_{xx'} \Big) \, ,
\end{eqnarray}
where 
\begin{eqnarray}
 & & W_{xx} \equiv w(x|x) := 0 \, , \nonumber \\
 & & w_x := \sum_{x'(\ne x)} w(x'|x) \, . 
\label{wxx}
\end{eqnarray}
$w_x$ is the total rate of transitions from state $x$ to all other
states.
\par
Applying the {\sc Laplace} transformation
\begin{equation}
 \tilde{f}(u) := \int\limits_{t_0}^\infty dt \; \mbox{e}^{-u(t-t_0)} f(t) 
\end{equation}
with a function, vector or matrix $f$ to equation (\ref{master2}) we obtain
\begin{equation}
 u \tilde{\bbox{p}}(u) - \bbox{p}(t_0) = (W - D) \tilde{\bbox{p}}(u) \, ,
\end{equation}
that means,
\begin{equation}
 \tilde{\bbox{p}}(u) = (u E - W + D )^{-1} \bbox{p}(t_0) \, ,
\end{equation}
where $E \equiv \Big( \delta_{xx'} \Big)$ is the unity matrix and $M^{-1}$
denotes the inverse of a matrix $M$. The desired solution of the master
equation (\ref{master2}) is, finally, 
found by inverse {\sc Laplace} transformation
\begin{equation}
 f(u) = \frac{1}{2\pi {\rm i}} \int\limits_{c-{\rm i}\infty}^{c+{\rm i}\infty}
 \!\! du \; \mbox{e}^{u(t-t_0)} \tilde{f}(u)
\end{equation}
of $\tilde{\bbox{p}}(u)$. It yields
\begin{eqnarray}
 \bbox{p}(t) &=& \frac{1}{2\pi {\rm i}} 
 \int\limits_{-c-{\rm i}\infty}^{-c+{\rm i}\infty}
 \!\! du \; \mbox{e}^{-u(t-t_0)} (D-uE-W)^{-1} \bbox{p}(t_0) \nonumber \\
 &=& \frac{1}{2\pi {\rm i}} 
 \int\limits_{-c-{\rm i}\infty}^{-c+{\rm i}\infty}
 \!\! du \; \mbox{e}^{-u(t-t_0)} [E - (D-uE)^{-1}W]^{-1} 
 (D-uE)^{-1} \bbox{p}(t_0) \, .
\label{invers}
\end{eqnarray}
The geometric series 
\begin{equation}
 [E - (D-uE)^{-1}W]^{-1} = \sum_{n=0}^{\infty} [(D-uE)^{-1}W]^n
\end{equation}
converges, if the constant $c$ is chosen sufficiently large. Taking into 
account
\begin{equation}
 [(D-uE)^{-1}W]_{xx'} = \sum_{x^{\prime\prime}} (w_x - u)^{-1} 
 \delta_{xx^{\prime\prime}} w(x^{\prime\prime}|x')
 = \frac{w(x|x')}{w_x - u}
\end{equation}
and (\ref{wxx}) we obtain
\begin{eqnarray}
& & \sum_{n=0}^\infty \sum_{x_0} \{ [ (D - uE)^{-1}W]^n (D-uE)^{-1}\}_{x_nx_0}
 p(x_0,t_0) \nonumber \\
&=& \sum_{n=0}^\infty \sum_{x_{n-1} (\ne x_n)} \dots \sum_{x_0(\ne x_1)}
\frac{w({\cal C}_n)}{\displaystyle \prod_{j=0}^n (w_{x_j} - u)}
 p(x_0,t_0)
\label{zwischen} 
\end{eqnarray}
with
\begin{equation}
 w({\cal C}_n) \equiv w(x_0\rightarrow \dots \rightarrow x_n) 
 := \left\{ 
\begin{array}{ll}
 \delta_{xx_0} & \mbox{if } n=0 \\
 \displaystyle \prod_{j=1}^n w(x_j|x_{j-1}) & \mbox{if } n \ge 1.
\end{array}\right.
\label{abk}
\end{equation}
Now, evaluating (\ref{invers}) by means of (\ref{zwischen}) 
and comparing the result
with (\ref{contraction}) leads to the explicit expression for the path
probabilities $p({\cal C}_n,\tau)$. It reads \cite{Hel93}
\begin{equation}
 p({\cal C}_n,\tau) = 
 \frac{1}{2\pi {\rm i}} \int\limits_{-c-{\rm i}\infty}^{-c+{\rm i}\infty}
 \!\! du \; \frac{\mbox{e}^{-u\tau} w({\cal C}_n)}
 {\displaystyle \prod_{j=0}^n (w_{x_j} - u)} p(x_0,t_0) \, .
\label{explicit}
\end{equation}
This important result
is in accordance with a detailed heuristic derivation of {\sc Empacher}
\cite{Nils}. By application of the residue theorem integral 
(\ref{explicit}) can easily be
evaluated. For example, if all total rates $w_x$ are identical (i.e.,
$w_x \equiv w$), we get \cite{Hel93}
\begin{equation}
 p({\cal C}_n,\tau) = \frac{\tau^n}{n!} 
 \mbox{e}^{-w\tau}w({\cal C}_n) p(x_0,t_0) \, ,
\label{exp1}
\end{equation}
whereas we obtain
\begin{equation}
 p({\cal C}_n,\tau) = \sum_{i=0}^n \frac{\mbox{e}^{-w_{x_i}\tau}w({\cal C}_n)}
 {\displaystyle \prod_{j=0 \atop (j\ne i)}^n (w_{x_j} - w_{x_i})}
 p(x_0,t_0) \, ,
\label{exp2}
\end{equation}
if the total rates $w_x$ are all different from each other.
Expressions (\ref{exp1}) and (\ref{exp2}) are polynomially increasing functions
in $\tau$ for $\tau \approx 0$ and exponentially decreasing functions for
$\tau \rightarrow \infty$.
\par
Note that for some cases the `path integral' (\ref{contraction}), 
(\ref{explicit}) can be analytically evaluated \cite{Nils}. Moreover,
a similar relation exists for the {\em generalized} master equation
\begin{equation}
 \frac{d\bbox{p}(t)}{dt} = \int\limits_{t_0}^t dt' \; 
 [W(t-t') - D(t-t')] \bbox{p}(t') 
\label{genma}
\end{equation}
with
\begin{equation}
 W_{xx'}(t-t') := \left\{
\begin{array}{ll}
 w(x|x';t-t') & \mbox{if } x' \ne x \\
 0 & \mbox{otherwise,}
\end{array}\right.
\end{equation}
since a {\sc Laplace} transformation of (\ref{genma})
leads to \cite{Kenkre}
\begin{equation}
 u \tilde{\bbox{p}}(u) - \bbox{p}(t_0) = [\tilde{W}(u) -
 \tilde{D}(u)] \tilde{\bbox{p}}(u) \, .
\end{equation}
In (\ref{abk}) and (\ref{explicit}) one has, therefore, only to replace
$w(x'|x)$ by $\tilde{w}(x'|x;u)$ and $w_x$ by $\displaystyle
\tilde{w}_x(u) := \sum_{x' (\ne x)} \tilde{w}(x'|x;u)$.

\section*{Path occurence times and computational methods}

In numerical evaluations of formula (\ref{contraction}), 
one can, of course, not sum up an infinite number of different paths.
One will rather restrict the summation to the paths which give the most
important contributions. In order to find out which paths are negligible,
we will calculate the probability distribution 
\begin{equation}
p(\tau|{\cal C}_n) := \frac{p({\cal C}_n,\tau)}{p({\cal C}_n)}
\end{equation}
of the occurence time $\tau$ of a path ${\cal C}_n$ given that 
this path is traversed. Here, we have introduced the normalization
factor
\begin{equation}
p({\cal C}_n) := \int\limits_0^\infty
 d\tau \; p({\cal C}_n,\tau) \, .
\end{equation}
We expect,
that $p(\tau|{\cal C}_n)$ is maximal for a certain time 
$\widehat{\tau}({\cal C}_n)$. On the one hand,
for $\tau \ll \widehat{\tau}({\cal C}_n)$ there is not enough
time for the $n$ transitions $x_{i-1} \rightarrow x_i$ to occur. On the
other hand, for $\tau \gg \widehat{\tau}({\cal C}_n)$ the likelihood of further 
transitions
\begin{equation}
 x_n \rightarrow x_{n+1} \rightarrow \dots \rightarrow x_{n+l}
\end{equation}
with $l\ge 1$ is great. In order to determine the mean 
$\langle \tau \rangle_{{\cal C}_n}$ of the {\em path occurence
time} $\tau$ and its variance
\begin{equation}
 \theta_{{\cal C}_n} := \langle (\tau - \langle \tau \rangle_{{\cal C}_n} )^2
 \rangle_{{\cal C}_n} = \langle \tau^2 \rangle_{{\cal C}_n}
 - ( \langle \tau \rangle_{{\cal C}_n})^2 
\end{equation}
we need the quantities
\begin{equation}
 \langle \tau^k \rangle_{{\cal C}_n} := \int\limits_0^\infty
 d\tau \; \tau^k p(\tau|{\cal C}_n) \, .
\end{equation}
By means of the residue theorem one obtains \cite{Hel93}
\begin{eqnarray}
 \langle \tau^k \rangle_{{\cal C}_n} &=& 
  \frac{1}{2\pi {\rm i} }
  \int\limits_{-c-{\rm i}\infty}^{-c+{\rm i}\infty} \!\! du \;
  \frac{w({\cal C}_n)}{\displaystyle p({\cal C}_n)
  \prod_{i=0}^n (w_{x_i} - u)} p(x_0,t_0) 
 \left(- \frac{d}{du} \right)^k \int\limits_0^\infty d\tau \;
  \mbox{e}^{-u\tau} \nonumber \\
 &=& \frac{w({\cal C}_n)}{p({\cal C}_n)}
  \frac{k!}{2\pi {\rm i} }
  \int\limits_{-c-{\rm i}\infty}^{-c+{\rm i}\infty} \!\! du \;
  \frac{p(x_0,t_0)}{\displaystyle 
  u^{k+1} \prod_{i=0}^n (w_{x_i} - u)} \nonumber \\
 &=& p(x_0,t_0) \frac{w({\cal C}_n)}{p({\cal C}_n)} 
 \left. \left( \frac{d}{du} \right)^k
 \prod_{i=0}^n \frac{1}{(w_{x_i} - u)} \right|_{u=0} \, . \nonumber \\
\end{eqnarray}
From this we can derive
\begin{equation}
 p({\cal C}_n) = p(x_0,t_0) w({\cal C}_n) \prod_{i=0}^n \frac{1}{w_{x_i}} \, ,
\end{equation}
\begin{equation}
 \langle \tau \rangle_{{\cal C}_n} = \sum_{i=0}^n \frac{1}{w_{x_i}} \, ,
\end{equation}
and
\begin{equation}
 \theta_{{\cal C}_n} = \sum_{i=0}^n \frac{1}{(w_{x_i})^2} \, .
\end{equation}
\par
In order to get an approximate solution of $p(x,t) = p(x,t_0+\tau)$, we have
only to sum up the path probabilities $p({\cal C}_n,\tau)$ of paths
${\cal C}_n$, for which
\begin{equation}
 | \tau - \langle \tau \rangle_{{\cal C}_n}| \le a \sqrt{\theta_{{\cal C}_n}
 \vphantom{N}}
\label{condi}
\end{equation}
holds. The parameter $a$ depends
on the desired accuracy of the approximation.
\par
In the following the computational solution method shall be explained in more
detail. Let us assume to have chosen $a=3$ for the accuracy parameter. Then,
about 99 percent of the distribution function $p(\tau|{\cal C}_n)$ are
found between $\tau_1 := \langle \tau \rangle_{{\cal C}_n} - a
\sqrt{\theta_{{\cal C}_n} \vphantom{N}}$ and $\tau_2 := 
\langle \tau \rangle_{{\cal C}_n} + a
\sqrt{\theta_{{\cal C}_n} \vphantom{N}}$. This implies that for every given
$\tau$ the paths ${\cal C}_n$ 
which fulfil condition (\ref{condi}) will
reconstruct about 99 percent of the distribution function $p(x,t_0+\tau)$
defined by (\ref{contraction}). Therefore, one possible method of numerically
calculating the distribution function $p(x,t_0+\tau)$ is the following:
\begin{itemize}
\item[1.] For every $x$ set $p(x,t_0+\tau) := 0$ and calculate
$w_x$ from $w(x'|x)$ according to (\ref{wxx}).
\item[2.] Let $n=0$ and generate all paths ${\cal C}_0 := x_0$ of
length $n=0$. Define the {\em relevance status} of each path ${\cal C}_0$
by setting $s({\cal C}_0):=1$. Calculate $\langle \tau \rangle_{{\cal C}_0}
:= 1/w_{x_0}$ and $\theta_{{\cal C}_0} := 1/(w_{x_0})^2$ for every path
${\cal C}_0$. If condition (\ref{condi}) is fulfilled, then add
$\exp (-w_{x_0}\tau) p(x_0,t_0)$ to $p(x_0,t_0+\tau)$. 
\item[3.] Let $n := n+1$ and generate all paths ${\cal C}_n
:= {\cal C}_{n-1} \rightarrow x_n$ for which the subpath ${\cal C}_n$
is relevant, that means, $s({\cal C}_n) = 1$. Calculate 
$\langle \tau \rangle_{{\cal C}_n} := \langle \tau \rangle_{{\cal C}_{n-1}}
+ 1/w_{x_n}$ and $\theta_{{\cal C}_n} := \theta_{{\cal C}_{n-1}} 
+ 1/(w_{x_n})^2$ for every generated path ${\cal C}_n$.
\item[4.] If condition (\ref{condi}) is satisfied, then calculate
$p({\cal C}_n,\tau)$ according to (\ref{exp2}) (or the corresponding
explicit formula of (\ref{explicit})) and add $p({\cal C}_n,\tau)$ to
$p(x_n,t_0+\tau)$. If $\tau < \langle \tau \rangle_{{\cal C}_n} - a
\sqrt{\theta_{{\cal C}_n} \vphantom{N}}$ is fulfilled, then cancel the
relevance of ${\cal C}_n$ by setting $s({\cal C}_n) := 0$.
\item[5.] If there are no relevant paths ${\cal C}_n$ with $s({\cal C}_n) = 1$
left, then stop the procedure. Otherwise procede with 3.
\end{itemize}
More refined, but more complicated algorithms 
can be developed for computers with little memory.

\section*{Conclusions and outlook}

We have derived a contracted and explicit path integral solution for the
discrete master equation. The result is also applicable to the generalized
master equation, and it allows the development of numerical methods of
solution.
\par
The advantage of the path integral solution with respect to other
solution methods is that one can calculate the probability
of a system to take `desired' paths or `catastrophic' paths consisting
of sequences of {\em selected} states only. This is of interest for 
prognoses (e.g., szenario techniques) or the (technical) control of a
system's temporal evolution, especially if the probability distribution
develops several maxima.

\section*{Acknowledgement}

The author wants to thank W. Weidlich and U. Wei\ss{}
for reading and commenting on the
manuscript.

\end{document}